\documentstyle[aps,epsfig]{revtex}
\begin{document}
\author{\footnotemark[1]\footnotemark[2]Jonathan A. Jones,
\footnotemark[1]Vlatko Vedral, \footnotemark[1]Artur Ekert and
\footnotemark[3]Giuseppe Castagnoli}
\title{Geometric Quantum Computation with NMR }
\address{\footnotemark[1]Centre for Quantum Computation, Clarendon
Laboratory, Parks Road, Oxford OX1 3PU, UK\\ \footnotemark[2]OCMS,
New Chemistry Laboratory, South Parks Road, Oxford OX1 3QT, UK\\
\footnotemark[3]Elsag,Via Puccini 2, 1615 Genova, Italy}
\date{\today}
\maketitle

\textbf{An exciting recent development has been the discovery that
the computational power of quantum computers exceeds that of
Turing machines \cite{Shor}.  The experimental realisation of the
basic constituents of quantum information processing devices,
namely fault-tolerant quantum logic gates, is a central issue.
This requires conditional quantum dynamics, in which one subsystem
undergoes a coherent evolution that depends on the quantum state
of another subsystem \cite{cqdlg95}. In particular, the subsystem
may acquire a conditional phase shift. Here we consider a novel
scenario in which this phase is of geometric rather than dynamical
origin \cite{MVB84,GPP89}.  As the conditional geometric (Berry)
phase depends only on the geometry of the path executed it is
resilient to certain types of errors, and offers the potential of
an intrinsically fault-tolerant way of performing quantum gates.
Nuclear Magnetic Resonance (NMR) has already been used to
demonstrate both simple quantum information processing
\cite{DGC97,GER97,Jones98,Chuang98b,DGC98} and Berry's phase
\cite{DS87,GFG96,JAJ97}. Here we report an NMR experiment which
implements a conditional Berry phase, and thus a controlled phase
shift gate.  This constitutes the first elementary geometric
quantum computation.}

Any quantum computation can be build out of simple operations
involving only one or two quantum bits (qubits)\cite{DD95}. A
particularly simple two qubit gate in many experimental
implementations, such as NMR \cite{Jones98c}, is the controlled
phase shift. This may be achieved using a conditional Berry phase,
and thus quantum geometrical phases can form the basis of quantum
computation. We will use spin half nuclei as an example to
demonstrate the feasibility of this approach, but the basic idea
is completely general. In our experiments the state of one spin
determines the Berry phase acquired by the other spin.

Suppose that a spin half nucleus undergoes a conical evolution
with cone angle $\theta$. Then the Berry phase is simply $\gamma
=\pm\frac{1}{2}\Omega=\pm\pi(1-\cos\theta)$ where the $\pm$ signs
depend on whether the system is in the eigenstate aligned with or
against the field, and $\Omega$ is the solid angle subtended by
the conical circuit. Note that any deformation of the path of the
spin which preserves this solid angle leaves the phase unchanged.
Thus the phase is not affected by the speed with which the path is
traversed; nor is it very sensitive to random fluctuations about
the path.

Berry phases can be conveniently demonstrated in an NMR experiment
\cite{Ernst87} by working in a rotating frame.  Consider an
ensemble of spin half particles in a magnetic field, $B_0$,
aligned along the $z$-axis; their precession frequency is then
given by the Larmor frequency, $\omega _{0}$.  If the spins are
irradiated by a circularly polarized RF field, $B_1$, at a
frequency $\omega _{\rm rf}$  the total Hamiltonian (neglecting
relaxation) may be written in the rotating frame as ${\cal
H}=\left(\omega_0-\omega_{\rm rf}\right)I_z + \omega_1 I_x$,
where, following conventional NMR practice, the Hamiltonian is
described in product operator notation \cite{OS83} and the field
strengths are written in terms of their corresponding Larmor
frequencies.

When $|\omega_1|\ll|\omega_0-\omega_{\rm rf}|$ the Hamiltonian
lies close to the $z$-axis, while when $|\omega_1| \gg
|\omega_0-\omega_{\rm rf}|$, the Hamiltonian lies close to the
$x$-axis. If RF radiation is applied far from resonance, the
system is effectively quantized along the $z$-axis, and if the RF
frequency is swept towards resonance ($\omega_{\rm rf}=\omega_0$),
the effective Hamiltonian rotates from the $z$-axis towards the
$x$-axis.  If the frequency sweep is applied adiabatically then
the spin will follow the Hamiltonian.  Next, a circular motion can
be imposed by adiabatically varying the phase of the RF. When the
Hamiltonian returns to the $x$-axis the frequency sweep may be
reversed, so that the spin returns to its original state, aligned
along the $z$-axis. The Berry phase acquired in this cyclic
process is $\pm\pi$. If the RF field is not swept all the way to
resonance, but only to some final value $\omega_f$, the
Hamiltonian ends at some angle to the $z$-axis, and so circuits
with arbitrary cone angles can be implemented.  A similar case
occurs if the frequency sweep is replaced by an amplitude sweep,
in which the RF is always applied away from resonance, and its
amplitude is raised smoothly from zero to some final value,
$\omega_1$.

This situation arises naturally in a system of two weakly coupled
spins, $I$ and $S$.  For simplicity we consider a heteronuclear
system, so that $\omega_I$ and $\omega_S$ are very different, and
only one spin (say $I$) is close to resonance. The two transitions
of $I$ (corresponding to the two possible states of spin $S$) will
be split by $\pm\pi J$, and so will have different resonance
offsets.  After an amplitude sweep the orientation of the
effective Hamiltonian depends on the resonance offset, and so
$\theta$ (and hence the Berry phase acquired) will depend on the
state of spin $S$. This permits a conditional Berry phase to be
applied to spin $I$, where the size of the phase shift is
controlled by spin $S$. If the RF is applied at a frequency
$\delta$ (measured in Hz) away from the transition frequency of
spin $I$ when spin $S$ (the control spin) is in state $0$, and
$\nu_1$ is the maximum RF field strength (also in Hz), then the
differential Berry phase shift,
\begin{displaymath}
\Delta\gamma=\gamma_1-\gamma_0=\pm\pi\left[
\frac{\delta+J}{\sqrt{(\delta+J)^2+\nu_1^2}}
-\frac{\delta}{\sqrt{\delta^2+\nu_1^2}} \right],
\end{displaymath}
depends only on $\delta$, $\nu_1$ and $J$; it is independent of
how the process is carried out as long as it is slow enough to be
adiabatic, but rapid compared with the decoherence times ($T_1$
and $T_2$).

In addition to the geometric phases, there will also be additional
dynamic phases, which \emph{do} depend on experimental details. In
principle these could be calculated and corrected for, but this is
not practical as a result of $B_1$ inhomogeneity. The RF field
strength will vary over the sample, and so different nuclei will
acquire different dynamic phases; averaging over the sample will
result in extensive dephasing. This can be overcome using a
conventional spin echo approach: the pulse sequence is applied
twice, with the second application surrounded by a pair of
$180^\circ$ pulses applied to spin $I$. This has the effect of
completely refocussing the dynamic phase, and thus refocussing any
inhomogeneity in it.  Note that this approach will only be
successful if the dynamic phase terms are the same during the two
halves of the spin echo, and thus it is important that any
variation in these terms occurs on a time scale long compared with
the echo time.  In our experiments minor variations in the dynamic
phase arising from the effects of molecular diffusion within the
slightly inhomogeneous $B_1$ field are visible as a small loss in
signal intensity.  Similarly it is important to ensure that
refocussing pulses are applied reliably, which is relatively
simple within NMR. Clearly these issues must be considered when
seeking to apply this approach with other experimental techniques.

This procedure would also cancel out the geometric phase, but this
can be side stepped by performing the RF phase sweep in the
\emph{opposite} direction, thus negating the geometric phase, so
that the two geometric terms add together while the dynamic phases
cancel out. Cancellation of dynamic phases arising from the
natural evolution of spin $S$, could also be achieved by
incorporating the sequence within another spin echo, involving
$180^\circ$ pulses applied to $S$.  Similarly, in more complex
spin systems, contributions to the geometric phase which depend on
the states of other spins can be cancelled by the judicious
application of further spin echoes.  It might seem that this
approach would require an exponentially large number of
refocussing pulses, but this is not true as nuclear spin--spin
coupling is a local effect, so that couplings between distant
nuclei can be safely neglected.  Furthermore it should be possible
to use efficient refocussing schemes\cite{Jones99} based on
Hadamard matrices, thus allowing the number of refocussing pulses
to be further reduced.

In order to measure the sizes of $\gamma_0$ and $\gamma_1$ it is
convenient to apply the Berry phase shifts to a spin $I$ in a
coherent superposition of states, created by an initial $90^\circ$
pulse.  The pulse sequence (figure~\ref{fig:pulses}) then
generates Berry phases that are determined by examining the final
phase of the magnetisation. As the 2 states of spin $I$ acquire
equal and opposite phases, and the pulse sequence contains two
separate periods in which phase shifts are generated, the total
phase change observed is $4\gamma$, with a maximum value of
$720^\circ$. A range of controlled Berry phases can be generated
by choosing appropriate values of $\delta$ and $\nu_1$ ($J$ is
fixed by the chemical system).  For a given value of $\delta/J$
the controlled phase will rise and then fall as $\nu_1$ is
increased. Clearly this approach will be most robust when the
desired $\Delta\gamma$ occurs at the maximum of this curve, as the
dependence on the size of $\delta$ is then reduced to second
order. For the particular case of a controlled $\pi$ shift, the
basis of the controlled-{\sc not} gate \cite{Jones98c}, this
occurs at $\delta = 1.058 J$ and $\nu_1 = 2.112J$.

NMR experiments were performed at $\rm25^\circ C$ using a
homebuilt $\rm500\,MHz$ ($\rm{}^{1}H$-frequency) NMR spectrometer
at the OCMS, with two power ranges for $\rm{}^{1}H$ pulses. Hard
pulses were applied using high power ($\rm\nu_1 < 25.8\,kHz$),
while adiabatic sweeps were performed using low power ($\rm\nu_1 <
774\,Hz$).  RF amplitude and phase calibration tables (available
on most modern NMR spectrometers) permit the RF power level to be
varied in a phase coherent manner.  The sample was prepared by
dissolving $\rm100\,mg$ of 99\% $\rm^{13}C$ labelled $\rm CHCl_3$
in $\rm0.2\,ml$ of 99.96\% $\rm CDCl_3$, and placing this in a
Shigemi microtube. The single $\rm{}^{1}H$ nucleus was used as
spin $I$, while the $\rm{}^{13}C$ nucleus was used as spin $S$;
for this system $J_{IS}=\rm209.2\,Hz$.  Spin--spin relaxation
times (measured using CPMG sequences and averaging over the two
components of each doublet) were $\rm3.9\,s$ for $\rm{}^{1}H$ and
$\rm0.3\,s$ for $\rm{}^{13}C$; the spin--lattice relaxation times
(measured by inversion-recovery) were $\rm7.6\,s$ for $\rm{}^{1}H$
and $\rm25.3\,s$ for $\rm{}^{13}C$.  Experiments were performed
with $\rm\delta=221.3\,Hz$, and $\nu_1$ was varied between zero
and its maximum value. Amplitude and phase sweeps were implemented
using 200 linear steps of $\rm 100\,\mu s$, giving a total pulse
sequence length of about $\rm 120\,ms$. The phases of the two
$\rm{}^{1}H$ resonances were determined by fitting the free
induction decay using home-written software. Reference phases were
obtained as described in the caption of figure~\ref{fig:pulses}.
The results are shown in figure~\ref{fig:results}; clearly the
measured phases lie close to the theoretically predicted values.
The controlled Berry phase rises smoothly to a broad maximum at
$180^\circ$, and then slowly falls back towards zero. In order to
investigate the effects of a breakdown in the adiabatic criterion,
some measurements were repeated with faster sweeps.  With a sweep
step size of $\rm50\,\mu s$ (data not shown) the results were
similar to, but not quite as good as, those obtained with the
slower sweep. Below $\rm50\,\mu s$ the loss of adiabaticity is
severe and major distortions are observed.  The step size should
not be increased too far beyond the adiabatic threshold, as this
will increase the effects of decoherence.

The conditional Berry phase gate demonstrated here depends only on
the geometry of the path, and is completely independent of how the
motion is performed as long as it is adiabatic; hence this kind of
computation may be called geometric quantum computation.  While
this approach has no particular advantage over more conventional
methods in NMR quantum computation, the basic idea is completely
general, and could be applied in other implementations. Some of
the methods described here have been partly foreshadowed in
previous theory papers (\emph{e.g.}, \cite{Pellizzari,Averin}),
but this is the first clear theoretical description and
experimental demonstration of such effects. This new approach to
quantum gates may be important in the future, as it is naturally
resilient to \emph{certain} types of errors. In particular,
suppose that the qubit, in addition to the circular motion which
implements the geometric phase, also undergoes a random motion
about its path due to some unwanted interaction with the
environment. Such noisy motion leaves the total area approximately
unchanged (although it changes details of the path), and so will
not be reflected in the final Berry phase. Geometric phases thus
offer the potential of performing quantum computations in a manner
which is naturally tolerant of some types of fault. Further
generalizations to non-abelian Berry phases, if implemented, may
open entirely new possibilities for robust quantum information
processing \cite{Kitaev,Preskill}.

\section*{Acknowledgements} Theory was developed by V.V, A.E., and
G.C.; NMR experiments were developed and performed by J.A.J\@. We
thank N. Soffe for helpful discussions. J.A.J. and A.E. are Royal
Society Research Fellows. J.A.J. and A.E. thank Starlab (Riverland
NV) for financial support. Correspondence should be addressed to
J.A.J. (e-mail: jonathan.jones@qubit.org).

\newpage
\begin{figure}
\vspace{1in}
\begin{picture}(220,40)\footnotesize
\put(0,20){\line(1,0){20}} \put(20,20){\line(4,3){20}}
\put(20,20){\line(4,-3){20}} \put(32,20){\makebox(0,0){$A_x$}}
\put(40,5){\framebox(20,30){}}
\put(50,20){\makebox(0,0){$\Phi_x$}} \put(60,35){\line(4,-3){20}}
\put(60,5){\line(4,3){20}} \put(68,20){\makebox(0,0){$\bar{A}_x$}}
\put(80,20){\line(1,0){18}} \put(98,5){\framebox(4,30){}}
\put(102,20){\line(1,0){18}} \put(120,20){\line(4,3){20}}
\put(120,20){\line(4,-3){20}} \put(132,20){\makebox(0,0){$A_x$}}
\put(140,5){\framebox(20,30){}}
\put(150,20){\makebox(0,0){$\bar{\Phi}_x$}}
\put(160,35){\line(4,-3){20}} \put(160,5){\line(4,3){20}}
\put(168,20){\makebox(0,0){$\bar{A}_x$}}
\put(180,20){\line(1,0){18}} \put(198,5){\framebox(4,30){}}
\put(202,20){\line(1,0){18}}
\end{picture}
\vspace{1in}\caption{Pulse sequence used to demonstrate controlled
Berry phases. Triangles indicate adiabatic RF amplitude sweeps,
from 0 to $\nu_1$ ($A_x$) or \emph{vice versa} ($\bar{A}_x$),
while rectangles indicate slow rotations of the RF phase at
constant amplitude; the phase rotation runs from $0$ to
$360^\circ$ ($\Phi_x$) or from $360$ to $0^\circ$
($\bar{\Phi}_x$). Narrow rectangles correspond to hard
$180^\circ_y$ pulses. As the absolute phase of an NMR signal is
undefined, it is essential to obtain a reference signal against
which experimental phases can be measured. The simplest approach
is to use the signal from a single $90^\circ$ pulse, while a more
subtle approach is to use this pulse sequence with the
$\bar{\Phi}_x$ sequence replaced by $\Phi_x$.  In principle these
should give the same result, but in practice minor differences are
seen as a result of RF inhomogeneity and the effects of the long
RF pulses on the NMR probe and pre-amplifier. } \label{fig:pulses}
\end{figure}
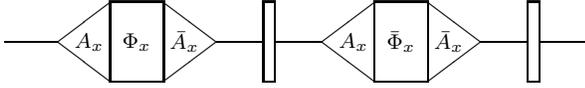
\begin{figure}
\vspace{1in} \psfig{file=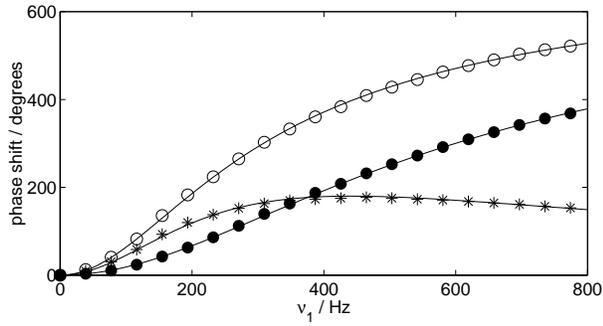,width=80mm} \vspace{1in}
\caption{Experimental results: $\gamma_0$, $\gamma_1$, and the
controlled Berry phase, $\Delta\gamma$, as a function of $\nu_1$.
Experimental points are shown by circles, the phase difference is
shown by stars, and theoretical values are shown as smooth curves.
Variability in the experimental points (estimated by repetition)
was about $\pm 2^\circ$; in a few cases the deviation of the
measured data points from their theoretical values was greater
than this, indicating the existence of as yet unidentified
systematic errors.  The signal strength observed after a phase
gate was about 90\% of that observed without a phase gate.  This
signal loss of about 10\% is too great to arise simply from
relaxation; more detailed experiments (data not shown) suggest
that the major source of signal loss is the effects of diffusion.
When considering the overall fidelity of the gate it is also
necessary to include effects arising from spin $S$; these are
dominated by the relatively rapid spin--spin ($T_2$) relaxation of
the $\rm{}^{13}C$ nucleus.} \label{fig:results}
\end{figure}
\end{document}